\documentclass[twocolumn,12pt]{article}
\usepackage{textcomp}
\usepackage{cite}
\usepackage{nicefrac}
\usepackage{amsmath}
\usepackage{upgreek}

\usepackage{graphicx}        % standard LaTeX graphics tool
\title{Surface Instabilities and Magnetic Soft Matter}
\author{Christian Gollwitzer,\\
{\sl Experimentalphysik V, Universit\"at Bayreuth, 95440 Bayreuth, Germany}
\and Marina Krekhova\footnote{present address:  Makromolekulare Chemie II,
Universit\"at Bayreuth, 95440 Bayreuth, Germany}, G\"unter Lattermann\footnote{present address:
Gr\"uner Baum 32, 95448 Bayreuth},\\
{\sl Makromolekulare Chemie I, Universit\"at Bayreuth, 95440 Bayreuth, Germany}\\
Ingo Rehberg and Reinhard Richter\\
{\sl Experimentalphysik V, Universit\"at Bayreuth, 95440 Bayreuth, Germany} }
\begin{document}

\maketitle
\hyphenation{extra-po-la-tion extra-po-late}
\noindent We report on the
formation of surface instabilities in a layer of thermoreversible ferrogel when
exposed to a vertical magnetic field. Both static and time dependent magnetic
fields are employed. Under variations of temperature, the viscoelastic
properties of our soft magnetic matter can be tuned. Stress relaxation
experiments unveil a stretched exponential scaling of the shear modulus, with an exponent 
of $\beta=1/3$. The resulting magnetic threshold for the formation of Rosensweig-cusps is measured
for different temperatures, and compared with theoretical predictions by
Bohlius \textit{et. al.} in {\em J. Phys.: Condens. Matter.}, 2006,
  {\bf 18}, 2671--2684.

\section{\label{subsec:elastic}Introduction}
Ferrofluids have found widespread applications in rotary seals, loud speakers,
or medicine \cite{berkovski1996,odenbach2009}. They are a colloidal dispersion
of magnetic nano particles in a carrier fluid like water or kerosene
\cite{rosensweig1985}. Due to their superparamagnetic behaviour, they can be
controlled by externally applied magnetic fields.

The same is true for ferrogels \cite{varga2003,collin2003,lattermann2006,zrinyi2007}, with the difference that here
the magnetic nano particles are embedded in an elastic (polymer) matrix.
This is handy for many upcoming applications, such as soft actuators, magnetic
valves, magnetoelastic mobile robots \cite{zimmermann2006}, artificial muscles
\cite{babincova2001} or magnetic controlled drug delivery \cite{lao2004,
francois2007,huang2007,liu2009}. Most of the applications are based on the
continuous \emph{deformation} of ferrogel bodies in gradient fields
\cite{szabo1998} or in homogeneous fields \cite{raikher2005b,gollwitzer2008}.

Another promising effect, which so far has not been exploited, is the drastic
shape transition of ferrogels in the succession of the Rosensweig instability.
If a threshold  of the magnetic induction is surpassed, the flat reference
state of the ferrogel is predicted to develop surface protuberances, pointing
in the direction of the vertically oriented field. The critical induction $B_c$
is derived by means of a linear stability analysis in Ref.\,\cite{bohlius2006}.
It depends on the surface tension $\sigma$, the mass density $\rho$, the gravity
$g_0$, and on the elastic shear modulus $G$ of the gel according to
\begin{equation}
B_c^2 = 2 \frac{\mu_0 \mu_r (\mu_r +1)}{(\mu_r-1)}(\sqrt{\sigma \rho g_0}+  G),
\label{eq:elastic.shift}
\end{equation}
where $\mu_0$ is the vacuum permeability and $\mu_r$ is the relative permeability of 
the ferrogel. 
In comparison to the Rosensweig instability in ferrofluids ($G=0$)
\cite{cowley1967} the threshold $B_c$ is enhanced by the elasticity due to an
increase of the surface stiffness. In contrast, the critical wavelength is
independent of the elastic modulus of the gel and is described by the capillary
wavelength
\begin{equation} k_c=\sqrt{ \frac{\rho g_0}{ \sigma}} \; ,
\label{eq:k.critical}
\end{equation}
familiar from ferrofluids \cite{cowley1967}. Like in the case of ferrofluids,
neither the critical field nor the critical wavelength depend on the viscosity.
Note that, in contrast to ferrogels, for the instability of a ferrofluid layer covered with a thin elastic film, 
it was predicted, that $k_c$ decays drastically with increasing elasticity~\cite{bashtovoi1978}.

More recently, the authors of Ref.~\cite{bohlius2006} derived the final pattern which forms in
ferrogels~\cite{bohlius2006b}. These results are limited to patterns of small
amplitude and were obtained via a minimisation of the energy density, similarly
to the method used for ferrofluids \cite{gailitis1977,friedrichs2001}. For
ferrofluids, the energy density comprises hydrostatic, magnetic and surface
terms. To account for the elastic deformations in ferrogels, an energy
contribution as given in \cite{jarkova2001} is added. The Rosensweig
instability in standard ferrofluids is associated with a transcritical
bifurcation and exhibits a hysteresis \cite{gailitis1977,bacri1984}. In
ferrogels, this hysteresis is expected to shrink  with increasing shear modulus
\cite{bohlius2006b}.

The Rosensweig instability in ferrofluids has been studied in many experiments;
see e.g. Refs.~\cite{cowley1967,browaeys1999,lange2000,richter2005,embs2007,groh2007,gollwitzer2007,richter2009}.
However, its counterpart in ferrogels is still awaiting measurements. The
reason for this void is that up to now mostly covalent cross linked polymer
gels \cite{zrinyi1996,collin2003,zrinyi2007} have been synthesised. This process
results usually in rather ``hard'' gels. Due to their high elasticity and the
saturation of the magnetisation, one can not excite surface instabilities in
these gels even for very high magnetic field strength. Only recently \emph{soft ferrogels} have been created
\cite{lattermann2006} which take advantage of thermoreversible, i.e. physically crosslinked gelators. 
Contrary to chemically crosslinked, irreversible ferrogels, their elasticity can be controlled by a thermoreversible sol-gel
transition. In the following we investigate the Rosensweig instability in such
a thermoreversible ferrogel.

%Selfassembly\\

%\section{Experimental}

\section{Material and Methods}
\subsection{Synthesis}
We prepared a thermoreversible ferrogel by swelling 5\,wt.\% of a gelator in an
oil-based ferrofluid containing 30\,wt.\% of magnetite particles. The carrier
liquid for the preparation of the ferrofluid was paraffin oil (Finavestan A50B
from Total Deutschland GmbH) with a kinematic viscosity
$\nu=13.6\,\mathrm{mm^2/s}$ at 20\,\textdegree C and a molar mass of
$280\,\mathrm{g/mol}$
(manufacturer information). The magnetite particles were prepared by
co-precipitation and stabilised with oleic acid~\cite{lattermann2006}. Transmission electron
micrographs show, that the diameter of the particles is $8\pm1\,\mathrm{nm}$~\cite{krekhova2008}. As a
gelator we have utilised Kraton~G\,1726 from Kraton Polymers, Belgium, which is
a mixture of 30\,wt.\% triblock copolymer and 70\,wt.\% of a diblock copolymer.
The triblock copolymer is poly(styrene-b-(ethylene-co-butylene)-b-styrene)
(SEBS) with a  molar mass of $\bar M_W =77\,700$ and a polydispersity index
$\text{PDI}=1.01$~\cite{krekhova2008}. The diblock copolymer (SEB) is exactly
one half of the triblock. The size distribution has been established by means of size
exclusion chromatography. For both the di- and triblock copolymer the styrene
content amounts to 30\,wt.\% (manufacturer information). In comparison with the
pure triblock copolymers used in earlier studies \cite{krekhova2008}, the
diblock acts here as a plasticiser and lowers the softening temperature to
25\,\textdegree C, according to the falling ball method \cite{lattermann2006}.
Both the ferrofluid and the ferrogel reveal no structure in optical
micrographs, i.e. they are perfectly homogeneous down to a sub-micrometer scale.
The ferrogel sample remains stable for 1\nicefrac{1}{2}\ year without any 
separation of the fluid phase from the gelator, i.e. without any syneresis.
Likewise, we did not observe changes of the magnetic properties and the
microstructure within this time.

\subsection{Material properties}
\label{sec:matprop}
\begin{figure}
\begin{center}
\includegraphics[width=1.0\columnwidth]{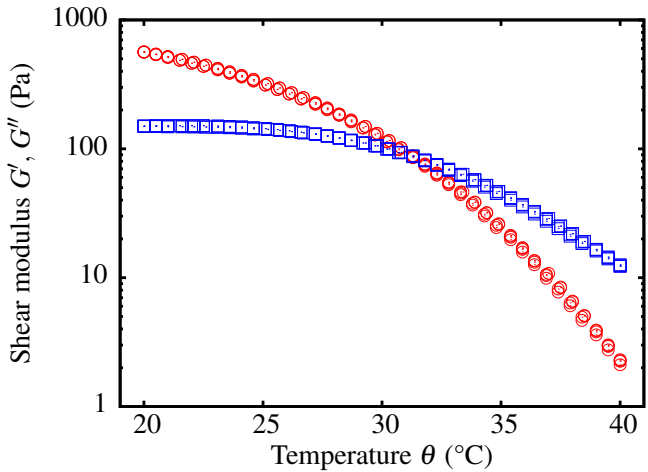}
\caption{Storage modulus $G^{\prime}$ (red circles) and loss modulus $G^{\prime
\prime}$ (blue squares) versus the temperature for the thermoreversible
ferrogel.} \label{fig:elastic.tempsweep}
\end{center}
\end{figure}

For the characterisation of the elastic properties we utilise a rheometer
(MCR\,301, Anton Paar) in cone-and-plate geometry. The cone has a diameter of
$50\,\mathrm{mm}$ and a base angle of $1$\textdegree. The rheometer is equipped
with a Peltier thermostated temperature device (C-PTD 200/E). Figure
\ref{fig:elastic.tempsweep} displays the result of oscillatory measurements of
the shear modulus at $1\,\mathrm{Hz}$ for a deformation of $\gamma=0.01$ versus the
temperature. $G^{\prime}$ and $G^{\prime \prime}$, which denote the real and
imaginary part of the shear modulus, respectively, have a crossover
around $31\,$\textdegree C. Above $\approx 45\,$\textdegree C, the sample becomes
liquid.

\begin{figure}
\begin{center}
\includegraphics[width=1.0\columnwidth]{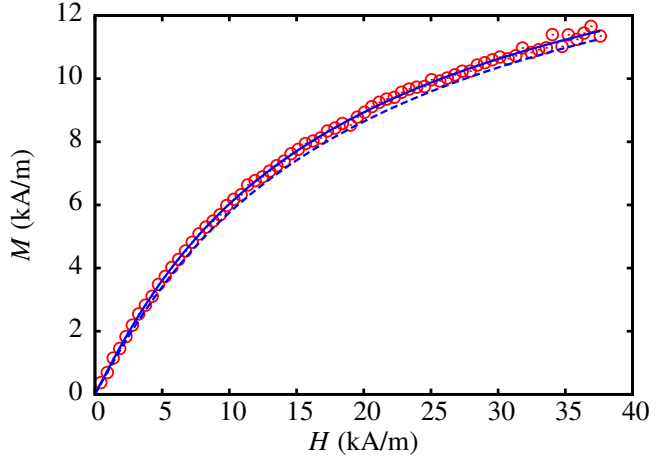}
\caption{Magnetisation as a function of the applied magnetic induction for the
thermoreversible ferrogel. The solid line represents a least squares approximation with the
model put forward by Ivanov \protect\cite{ivanov2001}. The 
best fit parameters for the gamma distribution are: 
exponent $a=8.1\pm1.3$, mean diameter $x_0\cdot(a+1)=8.1\pm0.4\,\mathrm{nm}$, volume fraction $\phi=3.9\pm0.09\,\%$. 
The core magnetisation has been held constant at $M_F=446\,\mathrm{kA/m}$.
The dashed line indicates the prediction for $\theta=38$\,\textdegree C with the above parameters.}
\label{fig:elastic.magnetization}
\end{center}
\end{figure}

%\subsection{Magnetic properties}
To measure the magnetisation curve of the ferrogel, 
we produce a sphere by casting $1.0\,\mathrm{ml}$ of the liquefied gel into a hollow
aluminium mold at $55\,$\textdegree C. For details see Ref.~\cite{gollwitzer2008}.
The magnetisation of the spherical sample was then measured by means of a
fluxmetric magnetometer (Lake shore, model 480) at $\theta=20\,$\textdegree C.
Figure~\ref{fig:elastic.magnetization} shows this data. 
The sample is superparamagnetic with an initial susceptibility of
$\chi_0=0.82$. This data has been fitted with the model of Ivanov
\cite{ivanov2001}, assuming a gamma distribution of the particle diameters. 
We use this model as well to extrapolate $M(H)$ for all sample temperatures, 
by evaluating the fit with a different $\theta$, while all other parameters are held constant.
As an example, the dashed line gives the extrapolation for the maximal applied temperature
of $\theta=38\,$\textdegree C.

The density $\rho$ of the ferrogel is
measured by immersing the sphere in water with added salt. The salt establishes
a vertical density gradient which keeps the ball floating in the middle of the container
at half depth.
With the help of a syringe, water has been sampled
directly above and below the ball. The density of both samples is then measured
by means of a vibrating tube density meter (DMA~4100, Anton Paar Co.). From the
mean of both values follows the density of the ball to be $\rho=1085 \pm
1\,\mathrm{kg/m^3}$. 

More cumbersome is an estimate for the surface tension $\sigma$.  As a rough
estimate, we measure the surface tension of the paraffin based ferrofluid the
gel was created from with a ring-tensiometer (LAUDA TE~2). We get $\sigma_\text{FF}=28.7\,\mathrm{mN/m}$ for both 
the ferrofluid and also the underlying paraffin oil that was used as a carrier liquid
for the ferrofluid. For obvious reasons, the tensiometer can not be directly applied to the ferrogel. 
Because the gelator is not surface active, it shall
not have a significant influence on $\sigma$. We therefore use $\sigma_\text{FF}$
in the subsequent calculations as the surface tension of the gel.
\begin{figure}
\begin{center}
\includegraphics[width=\columnwidth]{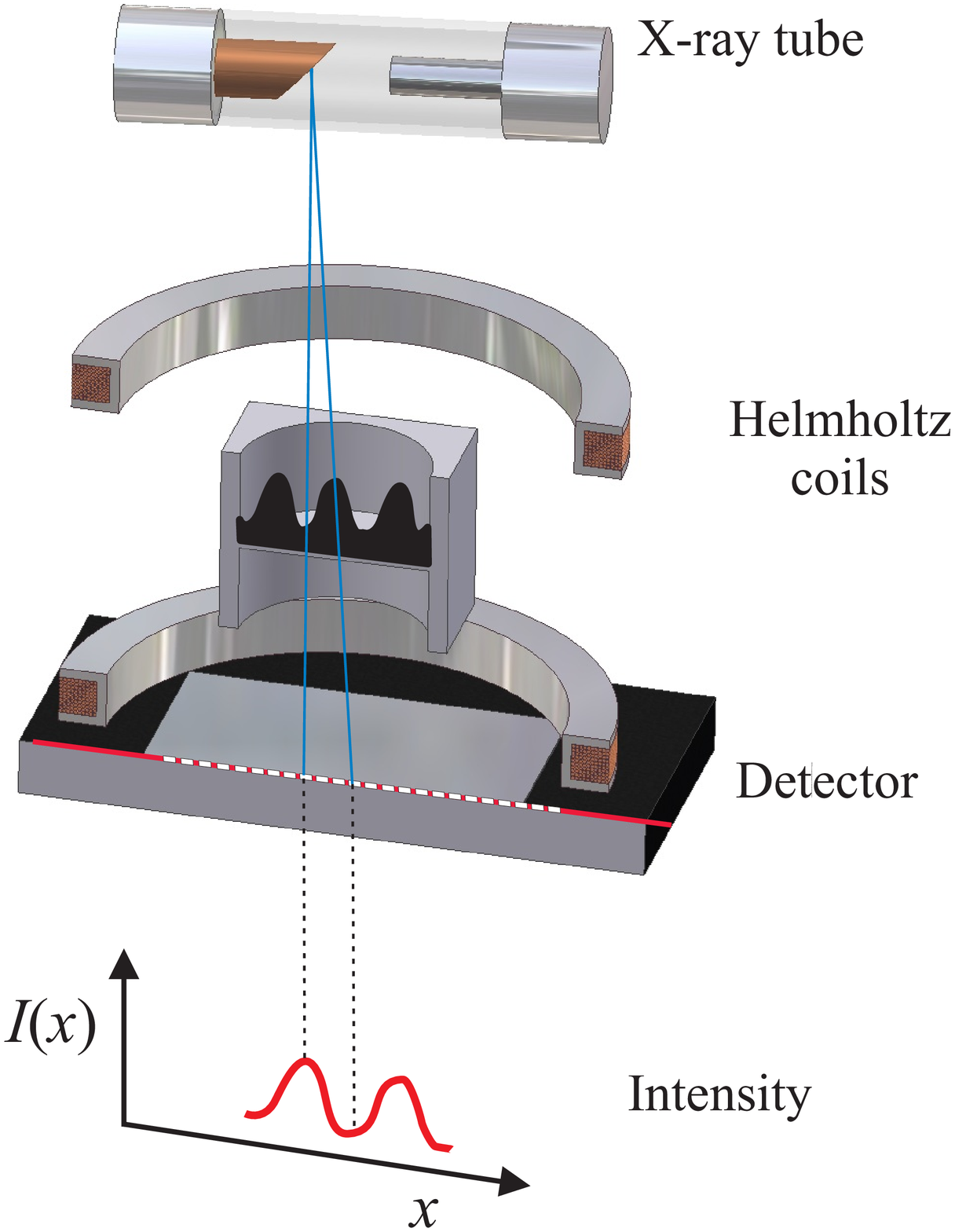}\\
(a)\\
\includegraphics[width=0.8\columnwidth]{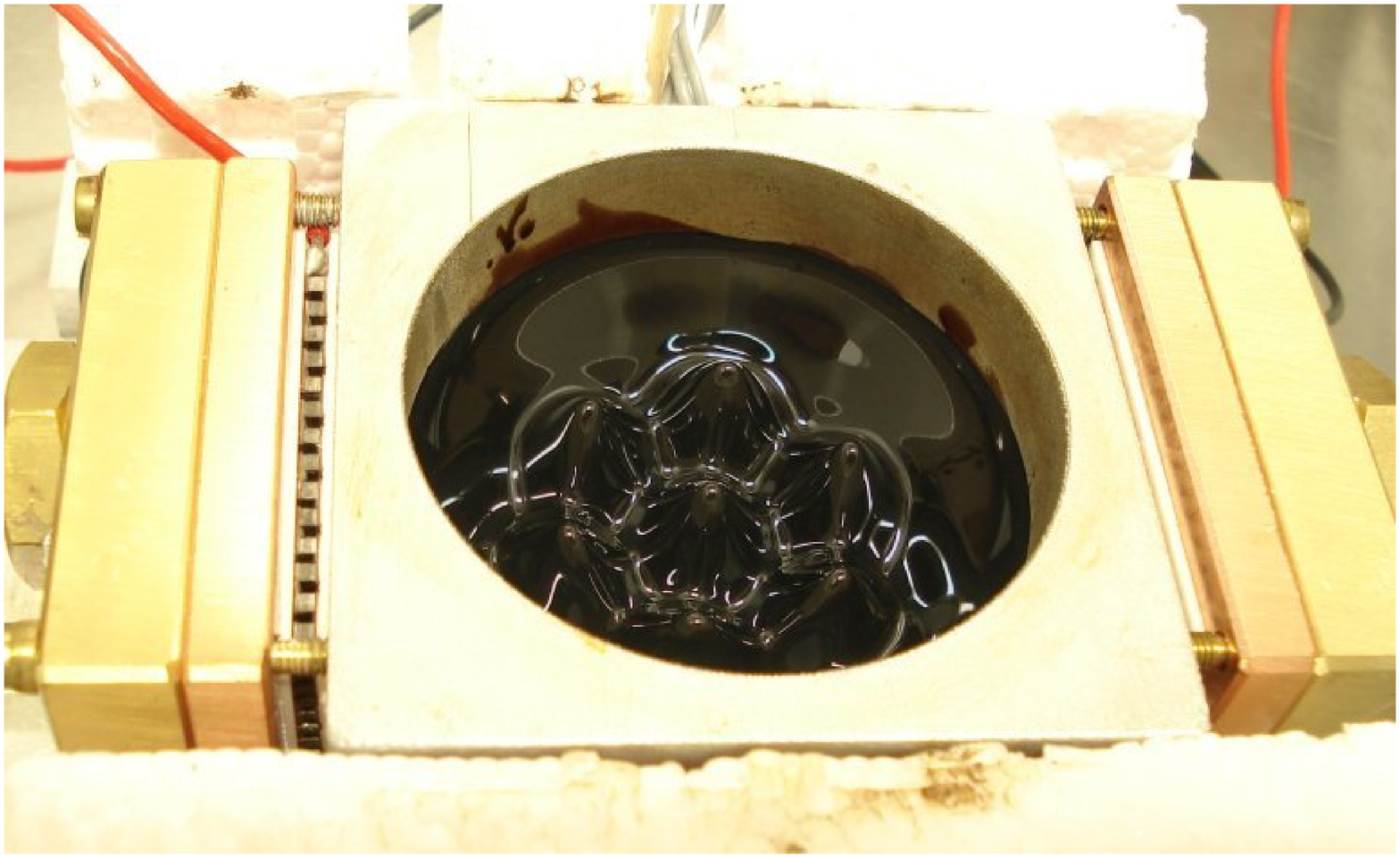}\\
(b)\\ \vspace{2mm}
\caption{Experimental setup: (a) Scheme of the setup (not to
scale); (b) photo of the temperature controlled container, filled with
ferrogel.} \label{fig:elastic.setup}
\end{center}
\end{figure}

\subsection{Setup for measurements of the normal field instability}
Our experimental setup is shown in Fig.~\ref{fig:elastic.setup}\,(a). In the
centre is a thermostated aluminium block of size (x,y,z)=(60,60,50)\,mm$^3$. A
cylindrical bore of $50\,\mathrm{mm}$ and a depth of $25\,\mathrm{mm}$ serves as a vessel for the
ferrogel. A bore with the same diameter, but a depth of $23\,\mathrm{mm}$ penetrates the
block from the lower side, in this way leaving a bottom plate with a thickness
of $2\,\mathrm{mm}$ (cf. Fig.~\ref{fig:elastic.setup}\,a). Utilising thermal grease, the
vessel is thermally connected to two Peltier elements, as shown in
Fig.~\ref{fig:elastic.setup}\,(b). They are equipped with heat exchangers
(1A~cooling Co., type 1A-SL2), which are circulated by water from a closed
cooling system (LAUDA RK20 KP). A thermo-resistor Pt100 serves to monitor the
temperature of the vessel. The Peltier elements are powered by a DC-current
source (EUROTEST Co., type LAB/SL 230/AI/LT) which is controlled via IEEE from
the computer. By a proportional-integral method, the computer regulates the
temperature of the vessel with a precision of $5\,\mathrm{mK}$ in the range of 
$-35\,$\textdegree C to $110\,$\textdegree C. The vessel is covered from above by means
of an aluminium plate with a thickness of $0.5\,\mathrm{mm}$. This lid creates an isolation
layer of air above the free surface of the ferrogel. We have chosen the small
aspect ratio $\Gamma=h/d\approx 5$ of the container, because the amount of
ferrogel was limited.

The vessel is fixed in the centre of a water cooled Helmholtz-pair-of coils
(Oswald Magnettechnik Co.). An X-ray tube is mounted above the centre of the
vessel at a distance of $1606\,\mathrm{mm}$. We measure the attenuation of X-rays passing
through the ferrogel layer in the vertical direction with a photodiode array
detector, which has a lateral resolution of $0.4\,\mathrm{mm}$. Rays which are
passing through crests are more attenuated than those passing through valleys
(cf. Fig~\ref{fig:elastic.setup}\,a). The full surface topography is then reconstructed from the radioscopic images. 
The dynamic range of the detector of $16\,$bits translates 
into a vertical resolution of $0.4\,\mathrm{\upmu m}$. This resolution is
achieved only when averaging many frames ($\approx 1000$), because of a noisefloor 
with an RMS value of $h_\text{RMS}=20\,\mathrm{\upmu m}$ for a single frame. 
The calibration of the absolute height is limited to $\approx 0.1\,\mathrm{mm}$ 
by the stability of the X-ray source and the mechanical positioning.
For details see Refs.  \cite{richter2001,gollwitzer2007}.

Prior to each series of measurements, a batch of $12\,\mathrm{ml}$ of ferrogel is positioned
in the empty vessel. This amount is molten by heating up the vessel to 
$90\,$\textdegree C, in this way creating a flat layer of ferrogel. Then the
temperature is lowered to the desired value for the measurement and held constant.

\section{Results}
%\subsection{Adiabatic Variation of the Induction}
\subsection{Quasistatic Experiments}
\begin{figure}
\begin{center}
\includegraphics[width=1.0\columnwidth]{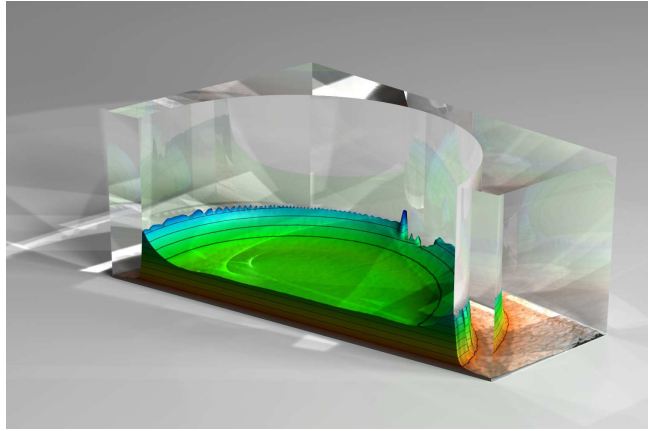}\\
(a)
\includegraphics[width=1.0\columnwidth]{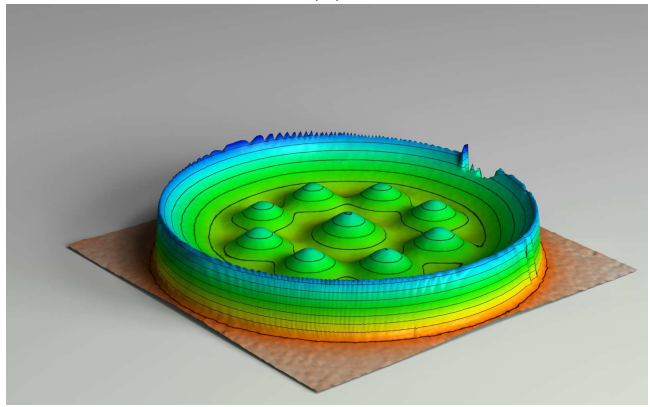}\\
(b) \caption{A Rosensweig pattern emerging in ferrogel, measured by radioscopy:
(a) cross section for $B=22.9\,\mathrm{mT}$; the transparent material illustrates the
size of the aluminium container; (b) full profile for
 $B=28.1\,\mathrm{mT}$. The black contour lines are indicating consecutive
 levels with a distance of $1\,\mathrm{mm}$.}
 \label{fig:elastic.relief}
\end{center}
\end{figure}

Figure \ref{fig:elastic.relief} presents characteristic topographies of the
ferrogel for subcritical (a) and supercritical inductions (b). Due to the
step-like jump of the magnetisation at the container edge, a field gradient arises which 
attracts the ferrogel towards the container edge. Therefore a meniscus is formed. Chart (b) shows an
example of the Rosensweig pattern. In the following, the height $h$ of the central peak of the
pattern serves as an order parameter and is estimated by fitting a paraboloid. 

\begin{figure}
\begin{center}
\begin{center}
\includegraphics[width=1.0\columnwidth]{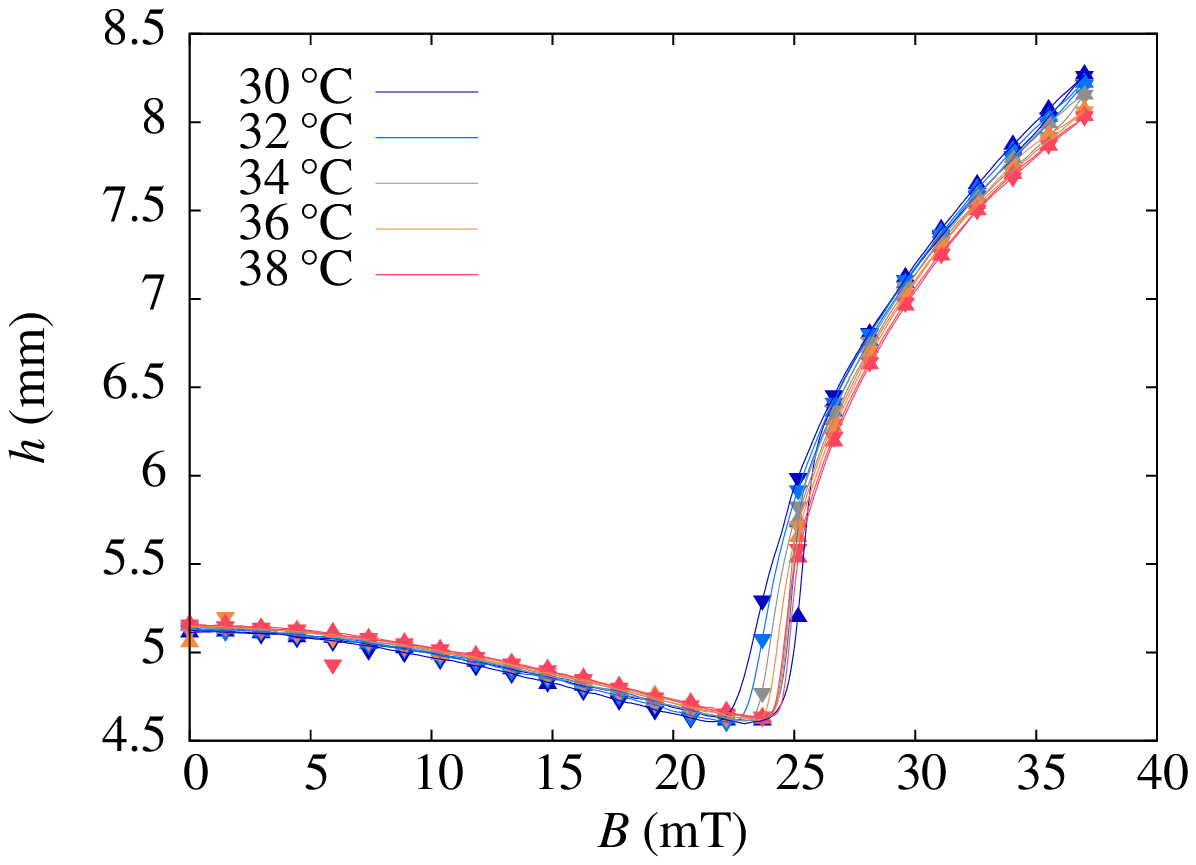}\\
(a)
\includegraphics[width=1.0\columnwidth]{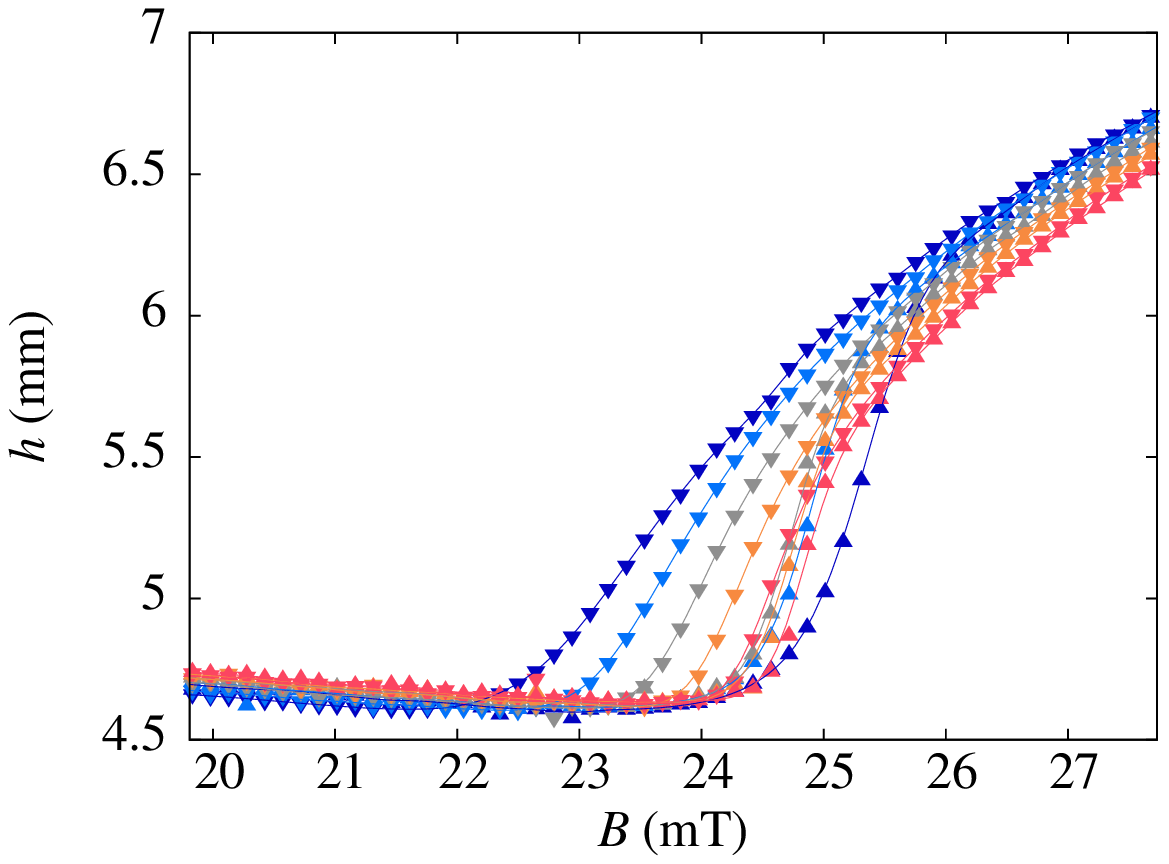}\\
(b)
\end{center}
\caption{Height of the central peak for an increase (upward triangles) and decrease
(downward triangles) of the magnetic induction. The colour encodes the temperature from
$30\,$\textdegree C~(blue) to $38\,$\textdegree C~(red), as in the legend. (a) Full range. For clarity
only every 10th data point is shown. The lines are splines to the full data
set. The time for the measurement was $2\,\mathrm{h}$. (b) Zoom of the hysteresis. All data
points are shown.} \label{fig:elastic.adiabatic}
\end{center}
\end{figure}

Figure~\ref{fig:elastic.adiabatic} gives the variation of the order parameter
for  a slow increase (upward triangles) or decrease (downward triangles) of the
magnetic induction for five different temperatures. In the interval from
$B=19\,\mathrm{mT}$ to around $24\,\mathrm{mT}$ one notices a monotonic decay. In this regime no
spikes exist. However, due to the formation of a meniscus (cf.
Fig.~\ref{fig:elastic.relief}\,a), the level of the material in the central
part of the vessel, where $h$ is estimated, decreases. For higher inductions, we
observe a steep increase of $h$ for all curves. It is this regime where the
ferrogel spikes are emerging.

Whereas in laterally infinitely extended liquid layers the transcritical bifurcation gives a proper scaling of the order parameter~\cite{friedrichs2001}, for small pools of ferrofluid, imperfections induced by the container edges obscure the analytical scaling law. In this case, only a numerical model is
available~\cite{spyropoulos2006,gollwitzer2009nfi}. In lack of an analytical
expression, we can not extrapolate the values for the critical induction $B_c$ from a fit of $h(B)$. As
an approximation for the threshold of the predicted discontinuous transition we
determine the induction $B_\textrm{max.up}$ where the amplitude $h(B)$ has its
steepest inclination, i.e. $\partial h/\partial B=\text{max}$. It is determined from a spline fit and listed in table
\ref{tab:elastic.matparam} for all investigated temperatures. A lower bound for
the threshold is given analogously from the data for decreasing magnetic induction. It is denoted by 
$B_\textrm{max.dn}$. The hysteresis, defined by $B_\textrm{max.up}-B_\textrm{max.dn}$, is in the
range of $2\,\mathrm{mT}$ at $30\,$\textdegree C and shrinks to a fraction of $1\,\mathrm{mT}$ at
$38\,$\textdegree C, as shown in the zoom presented in
Fig.~\ref{fig:elastic.adiabatic}\,(b).

\begin{figure}
\begin{center}
\includegraphics[width=1.0\columnwidth]{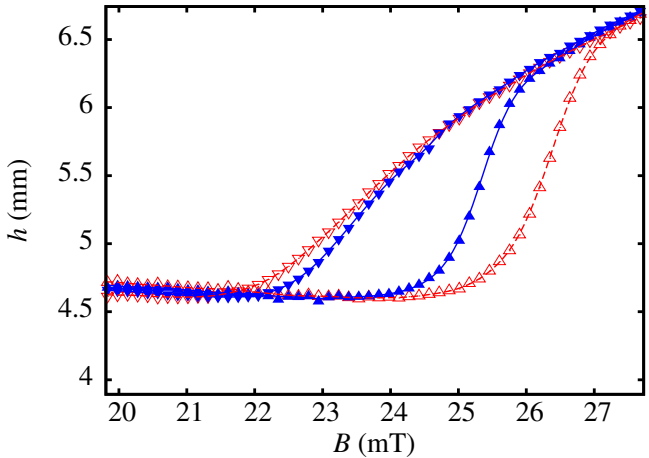}
\caption{Height of the central peak for increasing induction (upward triangles)
and for decreasing induction (downward triangles). The time between subsequent
data points was doubled from $14.4\,\mathrm{s}$ (red dashed line) to
$28.8\,\mathrm{s}$ (blue solid
line). The time for a full cycle from $0\,\mathrm{mT}$ to $37\,\mathrm{m}T$
amounts to $2\,\mathrm{h}$ and $4\,\mathrm{h}$,
respectively.} \label{fig:elastic.timetest}
\end{center}
\end{figure}

Next we check, whether the time for a measurement cycle in $B$ has an influence
on the evolution of the order parameter. In Fig.\,\ref{fig:elastic.timetest} we
present the results for a measurement cycle of $2\,\mathrm{h}$ (red dashed line)
and of $4\,\mathrm{h}$
(blue solid line) for a temperature of $\theta = 30$\,\textdegree C, which is
the lowest temperature, at which the measurements discussed beforehand have been
performed. One
clearly sees that the hysteresis between the upward- and downward branch
shrinks for longer cycle times. This variation is in the range of
$1\,\mathrm{mT}$. 

For higher temperatures (and lower $G$) the influence of the cycle time
becomes even less significant. Notwithstanding, the influence of the cycle time
indicates, that we have different time scales in the ferrogel. These time
scales will be studied next.

\subsection{Dynamic Experiments}
\subsubsection{Stress Relaxation Experiment}
\begin{figure}
\begin{center}
\includegraphics[width=1.0\columnwidth]{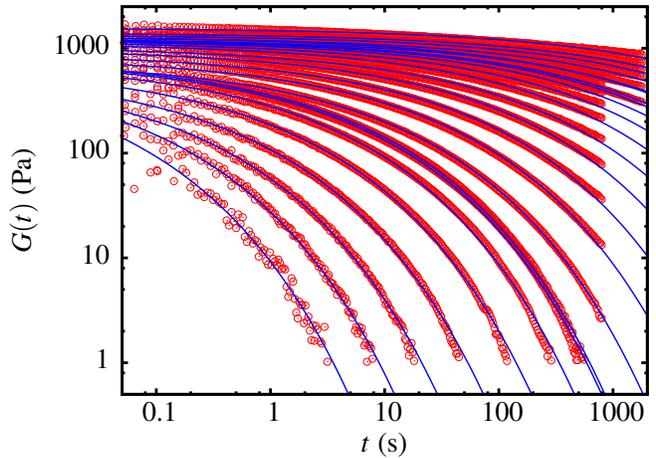}
\caption{Time dependent shear modulus $G(t,\theta)$ after a jump in the
deformation of $\gamma=0.01$ for different temperatures of the ferrogel. The
temperatures have been varied from $-10\,$\textdegree C to $32\,$\textdegree C.}
\label{fig:elastic.spann}
\end{center}
\end{figure}

In order to characterise the shear modulus $G$,
we perform a stress relaxation experiment for a series of temperatures. We
shear the sample by a deformation of $\gamma=0.01$ and measure the relaxation
of the stress $\tau$ while the deformation is held constant. Figure
\ref{fig:elastic.spann} displays the data points recorded for various
temperatures of the sample.

For a specific temperature, the restoring force decays drastically during one
second, which means that in our experiment, the shear modulus $G=\tau/\gamma$
cannot be treated as constant. 

Commonly, a stretched exponential function is used to describe the 
time dependent moduli of linear
viscoelastic media~\cite{kohlrausch1854,kohlrausch1863,williams1970,berry1997,anderssen2004} 
\begin{equation}
G(t,\theta) = G_0(\theta) \exp\left(
-\left(\frac{t}{t_0(\theta)}\right)^\beta\right) \label{eq:elastic.kww}.
\end{equation}
Here, the exponent $\beta$ is restricted to the range
$\left[0,1\right]$,
with $\beta=1$ for a simple exponential decay, and an increasingly broader
distribution of relaxation times for smaller values of $\beta$. 
Moreover, this scaling law was recently observed in the stress relaxation of a triblock copolymer subject to an extensional strain~\cite{hotta2002srt}, and consecutively explained in a model, 
which assumes that the copolymer reversibly splits into domains of different size~\cite{baeurle2005nsp}. These domains consist of a regular homogenous network of PS micelles which are interconnected by bridging chains of the middle block of the polymer. 
Also in our ferrogels, we observe glassy PS micelles arranged in clusters, which are interconnected by bridging chains of the ethylene-butylene middle block of the triblock copolymer gelator used. 
The size of these domains varies in the range from  $60$ to $120\,\mathrm{nm}$~\cite{lattermann2006,krekhova2008}.

Next, we apply~(\ref{eq:elastic.kww}) to our relaxation data. 
The solid lines in Fig.~\ref{fig:elastic.spann} give approximations of $G(t, \theta)$ with~(\ref{eq:elastic.kww})
for different temperatures $\theta$ from 
$\theta=32\,$\textdegree C down to $-10\,$\textdegree C. 
This temperature range was determined by the resolution of our rheometer.
All the fits use a common exponent $\beta$. The
best value amounts to $\beta=0.34 \pm 0.01$.

\begin{figure}
\begin{center}
\includegraphics[width=1.0\columnwidth]{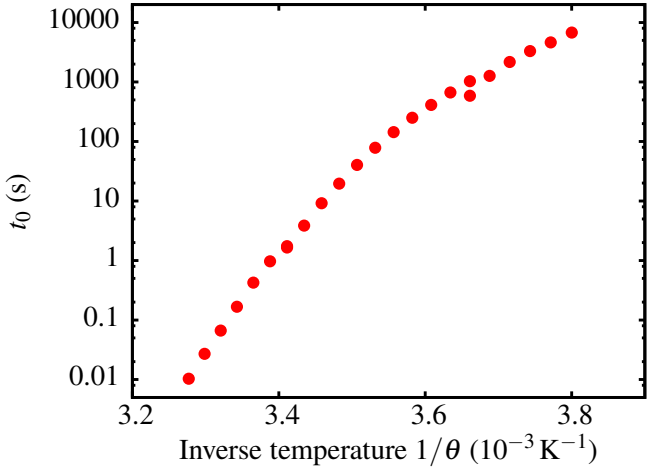}
\caption{Arrhenius plot of the characteristic relaxation time. The data stem from
fits of (\ref{eq:elastic.kww}) to the measurements presented in Fig.~\protect
\ref{fig:elastic.spann}.} \label{fig:elastic.taus}
\end{center}
\end{figure}

The characteristic relaxation time $t_0(\theta)$ of the ferrogel drops
drastically with increasing temperature. It varies over six orders of
magnitude, as shown in the Arrhenius plot Fig.~\ref{fig:elastic.taus}. 
However, the dependence does not make up a straight line -- therefore no simple 
Arrhenius behaviour with an activation energy can be inferred.  

Thus we observe for all temperatures a stretched exponential softening under
load with a characteristic scaling exponent near $\nicefrac{1}{3}$ for all
temperatures, while the characteristic relaxation time $t_0$ varies over six
decades. These experimental findings are in agreement with the observation of domains~\cite{krekhova2008} and
the predicted scaling law based upon a reversible splitting of domains~\cite{baeurle2005nsp}. 
%Remarkably, a constant value of $\beta=\nicefrac{1}{3}$ was predicted by
%de~Gennes \cite{gennes2002} for a polydisperse polymer melt 
%in the non-entangled limit. 
%

For $t\to\infty$, the time dependent shear modulus  $G(t,\theta)$ approaches
$0$. This means, that we have a viscoelastic soft magnetic material, without any
long term elasticity.

\subsubsection{Periodic Driving Magnetically}
In the effort of constraining $G$ to some finite value, 
%In search of a way out of the dilemma, 
we are next selecting a specific time scale.
We periodically drive the imposed magnetic induction
according to
\begin{equation}
B(t)=B_0+\Delta B \sin(2\pi f_D t). \label{eq:elastic.periodic.B}
\end{equation}
Here, $B_0$ denotes the bias value, $\Delta B=1.6\,\mathrm{mT}$ the driving amplitude,
and $f_D=1\,\mathrm{Hz}$ the driving frequency  of the induction imposed by the
Helmholtz-pair-of-coils. For each sample temperature, we measure the surface
response for 24 different values of $B_0$. For small $B_0$, the surface is
oscillating with $f_D$ around its meniscus-like shape, which becomes
alternatingly more and less pronounced concave. Beyond a threshold, spikes
appear which are oscillating with the amplitude $\Delta h$ around a mean height
$\bar{h}$ with the driving frequency. For the whole range of values, we observe
a harmonic response which can be described by
\begin{equation}
h(t)=\bar{h}+\Delta h \sin(2\pi f_D t + \phi). \label{eq:elastic.h}
\end{equation}

In order to determine the quantities $\bar{h}$ and $\Delta h$, we measure the
absorption of X-rays in the oscillating surface pattern by means of an X-ray
movie. For each data point we record 2200 frames with a frame rate of
$7.5\,\mathrm{Hz}$.
Each absorption picture $n$ is transformed into a height profile $h_n(x,y)$.
From the series of height profiles, we extract the desired quantities via the
equations
\begin{eqnarray*}
\bar{h}(x,y)  &=& \frac{1}{n} \sum_1^n h_n(x,y)
\label{eq:elastic.h.mean}\\
h_{\sin}(x,y) &=& \sum_1^n \sin (2\pi f_D n \Delta t)\,h_n(x,y)\,W(n \Delta t)
\label{eq:elastic.h.sin}\\
h_{\cos}(x,y) &=& \sum_1^n \cos (2\pi f_D n \Delta t)\,h_n(x,y)\,W(n \Delta t)
\label{eq:elastic.h.cos}\\
\Delta h(x,y) &=&
\sqrt{h_{\sin}^2(x,y)+h_{\cos}^2(x,y)}\,\mathrm{sgn}(h_{\sin})
\label{eq:elastic.h.delta},
\end{eqnarray*}\\
with
\begin{equation}
\mathrm{sgn}(x)= \left\{ \begin{array}{r@{\quad:\quad}l} -1 & x<0 \\ 0 & x=0 \\
+1 & x>0
\end{array} \right.
\end{equation}
The time delay between two consecutive frames amounts to $\Delta t=(1/7.5)\,\mathrm{s}$,
and $W(t)=N \exp\left(-(t-t_{1/2})^2/s^2\right)$ denotes a normalised
Gaussian window function with $s=0.4\,t_{1/2}$ and its centre at $t_{1/2}$, 
i.e. at the half of the measured time interval. 

\begin{figure}
\begin{center}
\includegraphics[width=1.0\columnwidth]{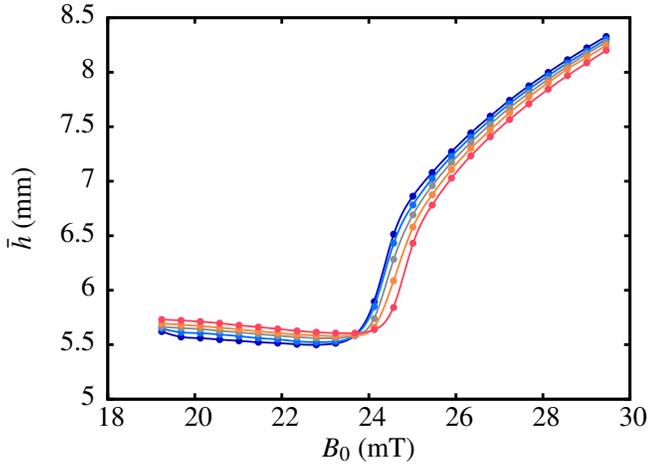}
\caption{Time averaged height $\bar{h}$ of the central peak of the surface
pattern (cf. Fig.~\protect \ref{fig:elastic.relief}) versus the bias of the
magnetic induction $B_0$. The lines stem from a spline fit.
The colour encodes the temperature as in Fig.~\protect\ref{fig:elastic.adiabatic}} \label{fig:elastic.bohlius.mittel}
\end{center}
\end{figure}

Figure \ref{fig:elastic.bohlius.mittel} displays the time averaged height of
the central extremum of the surface estimated in this way from the series of
measurements at five different temperatures. In the interval from
$\bar{B}=0\,\mathrm{mT}$ to around $24\,\mathrm{mT}$ one notices again a monotonic decrease of the
central height, due to the growth of the meniscus at the container edge. For
higher inductions, ferrogel spikes appear which again lead to a steep increase
of $\bar{h}$ for all curves.

Also here, we use the maximal inclination of $\bar{h}(B_0)$, which is determined from a
spline, as an estimate for the threshold. It is denoted by $\tilde{B}_{\rm
max.up}$ and is listed in table \ref{tab:elastic.matparam} for all 
temperatures investigated. This estimate for the threshold is shifted for higher
temperatures towards higher fields. The shift amounts only to
$\approx 0.5\,\mathrm{mT}$.

\begin{figure}
\centering
\includegraphics[width=0.9\columnwidth]{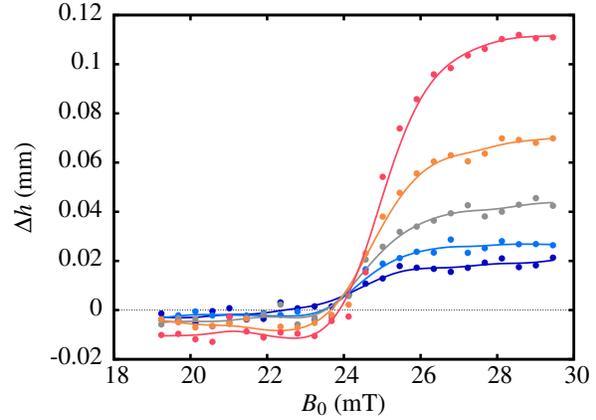}\\
(a)\\
\includegraphics[width=0.9\columnwidth]{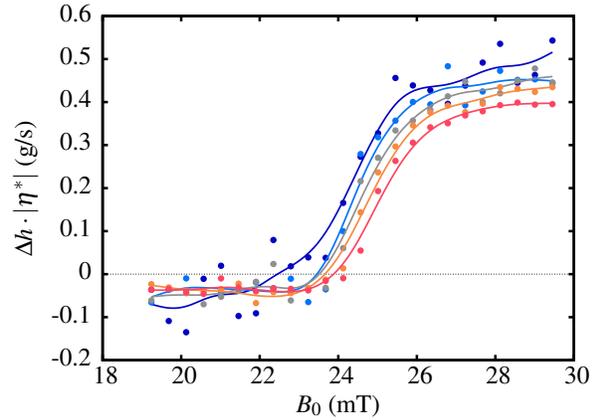}\\
(b)

\caption{Oscillation amplitude versus the bias of the magnetic
induction $B_0$.
(a) $\Delta h$ of the central peak (cf.Fig.~\protect \ref{fig:elastic.relief}) 
(b) Rescaled amplitude $\Delta h\cdot|\eta^*|$.
The lines stem from a spline fit. The colour code for the temperature is the same as in
Fig.~\protect\ref{fig:elastic.adiabatic}}
\label{fig:elastic.bohlius.real}
\end{figure}

Next, we present the oscillation amplitude $\Delta h$ of the central spike, as
shown in Fig.~\ref{fig:elastic.bohlius.real}\,(a). The amplitude is increasing with
increasing  temperatures, i.e. for softer ferrogel. From the
maximal increase, we determine an additional estimate for the threshold $\Delta
\bar{B}_\textrm{max.up}$. 
The results are listed in table \ref{tab:elastic.matparam}. Once
more the shift for different temperatures is only $\approx 0.5\,\mathrm{mT}$.

The X-ray movies of our oscillating spikes show, that the response is
always harmonic and never sub-harmonic. This is also true for a driven harmonic
oscillator. For strong damping and a driving frequency not much smaller than the resonant
frequency, the amplitude of an oscillator is inverse proportional to the damping constant.
To uncover such a scaling in our measurements, we plot
the rescaled amplitude $\Delta h\cdot|\eta^*|$ in Fig.~\ref{fig:elastic.bohlius.real}\,(b), 
where $|\eta^*|=|G|/\omega$ is the absolute value of the complex viscosity. Within the experimental resolution,
the graphs collapse onto a master curve. This indicates that the increase
of $\Delta h(B)$ under variation of $\theta$ stems solely from the softening of the ferrogel.
Therefore, one possible simplified explanation for our experiment could be an oscillator,
where the driving force comes from the magnetic stress at the edge of the
container and the viscosity provides the damping.

\subsection{Comparison of the thresholds with predictions}
Now we want to take a closer look at the experimentally determined thresholds
of the Rosensweig instability and how they compare to the predictions in Ref.~\cite{bohlius2006b}.
For a ferrofluid, the threshold can
be computed from the nonlinear magnetisation curve, the density $\rho$ and the
surface tension $\sigma$ via the linear stability analysis according to the book
by Rosensweig \cite{rosensweig1985}, \S~7.1. The critical magnetisation $M_c$
of the fluid layer is given by
\begin{equation}
M_c^2 = \frac{2}{\mu_0}\left(1 + \frac{1}{r_c}\right)\sqrt{g\rho\sigma}.
\label{eq:elastic.magkrit}
\end{equation}
Here $r_c = \sqrt{\mu_{ch}\mu_t}/\mu_0$ denotes the geometrical mean of the
chord permeability $\quad \mu_{ch} = \left.\frac{B}{H}\right|_{H_c}$ and the
tangent permeability \mbox{$\mu_t = \left.\frac{\partial B}{\partial H}
\right|_{H_c}$} at the critical field. Together with $M(H)$ and the jump
condition of the magnetic field at the base of the dish, $B=\mu_0 \left[H +
M(H)\right]$, the critical induction can be determined from these implicit
equations.

For a ferrogel with a Hookean shear modulus $G$ and a linear magnetisation curve, Ref.~\cite{bohlius2006b}
provides the expression (\ref{eq:elastic.shift}). 
We combine this equation with Eq.~(\ref{eq:elastic.magkrit}) 
for a non-linear $M(H)$ to get the more general form
\begin{equation}
M_c^2 = \frac{2}{\mu_0}\left(1 + \frac{1}{r_c}\right)(\sqrt{g\rho\sigma}+G).
\label{eq:elastic.magkrit.shift}
\end{equation}
If either one of the elasticity or non-linearity is left out, this equation reduces to
(\ref{eq:elastic.shift}) or (\ref{eq:elastic.magkrit}), respectively.

In table~\ref{tab:elastic.matparam}, we present the calculated critical inductions $B_\textrm{cFG}$ 
according to Eq.~(\ref{eq:elastic.magkrit.shift}). For
all calculations, we utilise the storage modulus $G^{\prime} (1\,\mathrm{Hz})$ as
determined by the oscillatory measurements (cf. Fig.~\ref{fig:elastic.tempsweep}).
The other properties entering Eq.~(\ref{eq:elastic.magkrit.shift}) are taken from \S~\ref{sec:matprop}, accordingly.
Specifically the surface tension is not well known for our gel. The value used here
was $\sigma_\text{FF}=28.7\,\mathrm{mN/m}$. Calculations for $\sigma=35\,\mathrm{mN/m}$ 
show that a variation of $\sigma$ does not change $B_\text{cFG}$ by more than $5\,$\%.
The magnetisation curve has been modeled by the equations given in~\cite{ivanov2001}. 

The comparison between the experimentally determined thresholds in table~\ref{tab:elastic.matparam} and 
the predictions in Ref.~\cite{bohlius2006b} reveals two prominent differences. 

\begin{figure}
\begin{center}
\includegraphics[width=\columnwidth]{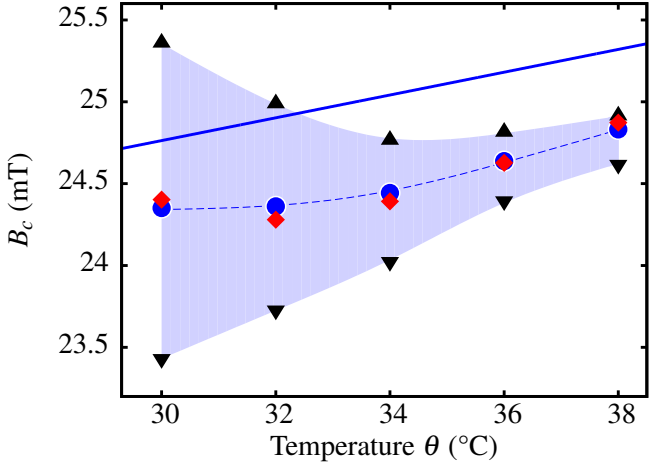}
\caption{Temperature dependence of the thresholds. 
The black upward (downward) triangles mark
$B_\textrm{max.up}$ ($B_\textrm{max.dn}$) for an adiabatic increase (decrease),
respectively. The shaded area in between denotes the hysteresis. The blue circles
and red diamonds show $\tilde{B}_\textrm{max.up}$ ($\Delta \tilde{B}_\textrm{max.up}$), respectively. The dashed line is just a guide for the eye. 
The theoretical value for a ferrofluid is displayed by the solid line.}
\label{fig:elastic.Bc.lin}
\end{center}
\end{figure}

\emph{Firstly,} we observe a decrease of the hysteresis of the pattern amplitude for a decrease of $G$,
i.e. higher temperatures, when the magnetic induction is varied in a quasi-static manner.
This hysteresis is denoted in Fig.~\ref{fig:elastic.Bc.lin} by the shaded area 
and the upward and downward oriented full triangles.
In contrast, Ref.~\cite{bohlius2006b} predicts an increase of
the hysteresis under reduction of $G$ for a ferrogel with \emph{Hookean
elasticity}.

This can be explained from the temperature dependent relaxation process of the
ferrogel.  The relaxation times are increasing from
$\tau \approx 0.01\,\mathrm{s}$ (at $32\,$\textdegree C) to $\tau \approx
10000\,\mathrm{s}$ (at $-10\,$\textdegree C). At the same time the measurement protocol for a full
ramping of the magnetic induction was kept at $2\,\mathrm{h}$. The hysteresis increases
because the material needs more and more time to follow a variation of $B$. We
have checked that by increasing the cycle time from $2\,\mathrm{h}$ to
$4\,\mathrm{h}$. The hysteresis was diminished, as shown in Fig.~\ref{fig:elastic.timetest}.

\emph{Secondly,} whereas $G^\prime(1\,\mathrm{Hz})$ varies in the investigated temperature
range over two decades (cf. Fig.\,\ref{fig:elastic.tempsweep}), the threshold
of the instability is varying only within 10\,\%. This is in contrast to the
model \cite{bohlius2006b} for a ferrogel with
\emph{Hookean elasticity}. However we found a different mechanical behaviour of our
material. The shear modulus is strongly time-dependent and even vanishes for
$t\to\infty$. 

\begin{figure}
\begin{center}
\includegraphics[width=\columnwidth]{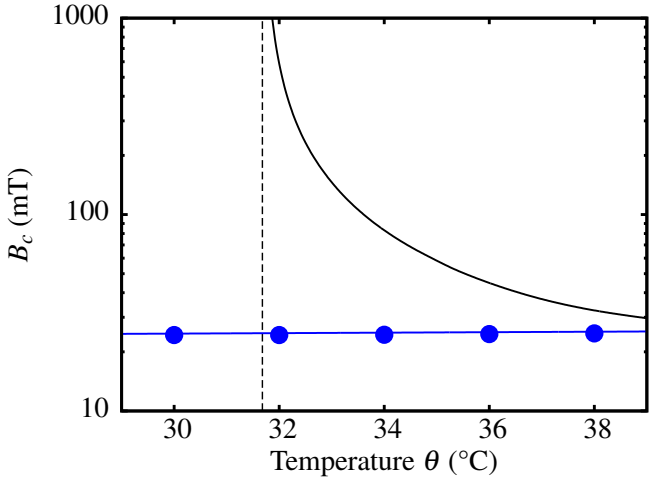}
\caption{Comparison of theoretically (solid black line) and experimentally (filled circles) determined critical inductions, as listed in table \protect \ref{tab:elastic.matparam}. The solid blue line is the theory for a ferrofluid. Note, that in this scale, all experimental values collapse onto that line. The dashed line denotes the vertical asymptote of the theoretical curve at $\theta=31.7\,$\textdegree C.} \label{fig:elastic.Bc}
\end{center}
\end{figure}

In an attempt to select a specific, finite value for $G$, we modulated the
magnetic induction with a fixed amplitude and a frequency of $1\,\mathrm{Hz}$ and applied
this magnetic driving together with a magnetic bias induction $B_0$. Under
increase of $B_0$ a threshold is overcome and we observe a steep increase of
the time averaged peak height $\bar{h}$. The magnetic induction at which this
transition occurs is marked in Fig.\,\ref{fig:elastic.Bc.lin} by the blue
filled circles. Also the oscillation amplitude $\Delta h$ shows a steep
increase under variation of $B_0$. The corresponding thresholds are shown in
Fig.~\ref{fig:elastic.Bc.lin} by red diamonds. Within the experimental scatter
both thresholds coincide. Similar to the static experiment, their values vary
only in fractions of $1\,\mathrm{mT}$ under decrease of the sample temperature $\theta$.

In contrast, the theoretical value, estimated according to
(\ref{eq:elastic.magkrit.shift}) at $G^\prime(1\,\mathrm{Hz})$, increases drastically under
decrease of $\theta$, as shown in Fig.\ref{fig:elastic.Bc}. Only for high
temperatures, the experimental and theoretical values are close to each other.
For lower temperatures, the gap between them increases drastically.
According to (\ref{eq:elastic.magkrit.shift}) it even diverges at a critical
shear modulus of
\begin{equation}
G_\mathrm{c} = \frac{\mu_0}{4} M_\mathrm{S}^2 - \sqrt{\rho g_0 \sigma} = 77.8\,
\rm Pa. \label{eq:elastic_Gc}
\end{equation}
For our gel, the saturation magnetisation $M_S$ of the ferrogel amounts to
$M_\mathrm{S}=14.7\,\mathrm{kA/m}$ and the divergence occurs at 
$\theta=31.7\,$\textdegree C. Obviously, there is a mismatch of our complex, soft material and
the linear model (\ref{eq:elastic.magkrit.shift}) when implementing 
$G^\prime(1\,\mathrm{Hz})$.

Let us look more closely on the small variation of the experimentally
determined thresholds in Fig.~\ref{fig:elastic.Bc.lin}. We see that under an
increase of the temperature all estimates for the onset of the instability are
slightly shifted to higher inductions. This shift can be understood
from the fact that for increasing $\theta$ the magnetisation diminishes.
Utilising the $M(H,\theta)$-model by Ivanov \emph{et al.} \cite{ivanov2001}, we
have taken into account this effect. The corresponding results for $B_\textrm{cFF}$ are
marked in Fig.\,\ref{fig:elastic.Bc.lin} by the solid line. The good agreement
between the experiment and the plain Rosensweig estimate for the threshold computed
in this way indicates that the shear modulus $G^\prime(1\,\mathrm{Hz})$ has no influence on the
threshold at all.

\section{Conclusion}
To conclude, we conducted the first measurements of the Rosensweig instability
in thermoreversible ferrogels. The present material shows a complex viscoelastic relaxation
process with an interesting critical exponent $\beta\approx\nicefrac{1}{3}$,
for experiments carried out in the time domain. It is possibly explained by a reversible splitting 
of the polymer network into domains of different size~\cite{lattermann2006,baeurle2005nsp}.  
Due to this relaxation, the threshold of the Rosensweig peaks is not much different from the 
Rosensweig instability in ferrofluids under an adiabatic increase of a
static magnetic field. However, the time scales are much
slower. This is especially pronounced for the lower part of the investigated
temperature range. Experiments with a periodic modulation of the magnetic field show that the 
\emph{complex viscosity} can be used to describe the response of the ferrogel. 
 
Such a complex elastic behaviour, however, is unsuitable to proof or rebut the
model of the Rosensweig instability~\cite{bohlius2006b}, which was derived for just a Hookean shear
modulus. Certainly soft matter with finite $G$ for $t \rightarrow \infty$ would
come closer to that model. One may think that a ferrogel with a
higher gelator concentration will serve this aim. However, as a consequence for
such a gel, the instability is completely hindered by the strong elastic
modulus. Unfortunately, the amount of magnetite cannot be increased further to 
overcome the critical magnetisation~(\ref{eq:elastic.magkrit}). 
Here, magnetic gels incorporating particles with a higher saturation magnetisation
could solve this problem. Despite some attempts, with e.g. cobalt particles,
such gels are not yet available.

\subsection*{Acknowledgements}
The authors thank H.~R.~Brand, H.~Schmalz, and E.M.~Terentjev for helpful discussions and 
K.~Oetter for constructions. We are grateful to Martina Heider for measurements 
of the surface tension. 
Financial support by \emph{Deutsche
Forschungsgemeinschaft} under the cooperative research project \emph{surface
instabilities in ferrogels} within research group FOR\,608 (nonlinear dynamics of
complex continuous matter) is gratefully acknowledged.

\clearpage

\bibliographystyle{pccp} %apsrev  unsrt
\bibliography{abspeck}

\clearpage
\onecolumn
The tables should be submitted normally after the
reference list, starting on a separate page.

\begin{table}[htbp]
\caption{Critical inductions for various temperatures: $B_\textrm{cFF}$ gives the
results of (\ref{eq:elastic.magkrit.shift}) for $G=0\,\mathrm{Pa}$, i.e. for a virtual
ferrofluid with otherwise the material parameters of the ferrogel. $B_\textrm{cFG}$
(mT) is estimated from $G^\prime$ at $1\,\mathrm{Hz}$ and the measured value of $\sigma$.
Experimentally, the thresholds were determined from the maximal incliniation of $h(B)$, 
as marked by $B_\textrm{max.up}$ for increasing $B$ and $B_\textrm{max.dn}$ for decreasing $B$.
From the oscillatory measurements we determine the
maximal inclination of $\bar{h}(B)$, marked by $\tilde{B}_\textrm{max.up}$, and the
maximal inclination of $\Delta h(B)$ abbreviated by $\Delta \tilde{B}_\textrm{max.up}$}
\begin{center}
\begin{tabular}{c|cc|cc|cc}
\hline \hline \rule{0.0em}{1.1em}
$\theta$       & $B_\textrm{cFF}$ & $B_\textrm{cFG}$ &
$B_\textrm{max.up}$ & $B_\textrm{max.dn}$ &$\tilde{B}_\textrm{max.up}$&$\Delta
\tilde{B}_\textrm{max.up}$ \\
(\textdegree C)&  (mT)     & (mT)   & (mT) &  (mT)  & (mT)  & (mT)\\
\hline
30 & 24.76     & $\infty$ & 25.36 & 23.43 & 24.35 & 24.40 \\
32 & 24.90     & 587.09   & 24.99 & 23.73 & 24.36 & 24.28 \\
34 & 25.04     & 83.17    & 24.76 & 24.02 & 24.44 & 24.39 \\
36 & 25.18     & 32.51    & 24.81 & 24.39 & 24.64 & 24.63 \\
38 & 25.32     & 28.03    & 24.91 & 24.62 & 24.83 & 24.87 \\
\hline \hline
\end{tabular}
\end{center}
\label{tab:elastic.matparam}
\end{table}

%30 22.986
%32 23.2084
%34 23.4309
%36 23.7067
%38 23.6533

%30 25.3587 23.4309
%32 24.988 23.7275
%34 24.7655 24.024
%36 24.8145 24.3948
%38 24.9138 24.6172

%Please compile a list of all figure captions on a separate page:
\clearpage

\listoffigures

\clearpage

\end{document}